\documentclass{article}
\usepackage{arxiv}
\usepackage[utf8]{inputenc} 
\usepackage[T1]{fontenc}    
\usepackage{hyperref}       
\usepackage{url}            
\usepackage{booktabs}       
\usepackage{amsfonts}       
\usepackage{nicefrac}       
\usepackage{microtype}      
\usepackage{lipsum}
\usepackage{amsmath,amssymb,graphicx}

\title{An exact solution of the orbit equation for a massive particle in Schwarzschild metric}

\author{
  A. S. Ribeiro\\
  GTMCOQ, Instituto Federal do Piau\'i -- IFPI \\
  São Raimundo Nonato, Piau\'i, 64770-000, Brazil.\\
  and\\
  Departamento de F\'{\i}sica, Universidade Federal do Cear\'{a}, Caixa Postal 6030, Campus do Pici,\\ 60455-760, Fortaleza, Cear\'{a}, Brazil.\\
   \texttt{antonius.ribeirus@ifpi.edu.br} \\
   \And
 F. N. Lima \thanks{I am corresponding author.} \\
  GTMCOQ, Instituto Federal do Piau\'i -- IFPI \\
  São Raimundo Nonato, Piau\'i, 64770-000, Brazil.\\
   \texttt{nogueira@ifpi.edu.br} \\
  }

\begin{document}
\maketitle

\begin{abstract}
In this paper, we consider a spherically curved symmetric spacetime to exact solving the orbit equation of a massive particle by using Jacobi's elliptic functions. Generally, the solution of the orbit equation provides the relativistic effects on the massive particle, absents in Newtonian mechanics. Besides, we investigate the additional physical information introduced by the exact solution to the orbit equation that is not visible in the approximate solutions traditionally presented in literature. Here, we  exactly solve the problem by the use an analytical methodology step by step in order to provide detailed solutions as well as demonstrate with mathematical rigour the geodesic solution in terms of Jacobi's elliptic functions.  We find oscillatory movements of the orbit of the massive particle at the expected regimes without to consider any heuristic argument. Bound regions to the solution of the equation of motion is presented, finding the aspect of the geodesic when the massive particle is trapped in the gravitational field of the source.
\end{abstract}

\keywords{Field theories in dimensions other than four \and Relativity and gravitation \and Self-gravitating systems; continuous media and classical fields in curved spacetime \and Special functions.}

\section{Introduction}
\label{section1}
\qquad 
Since the seminal work of Einstein in 1915 on general relativity \cite{Einstein:1915ea}, we have know this theory is the basis to appropriately describe many problems considering a real fact of nature, i. e., the gravity is nothing but the curvature of spacetime. The great success of the theory is related to the several experiments through history that clearly proven the ideas introduced by Einstein in his theory. As examples of those experiments, we have the deviation of light rays due to the curvature of the spacetime, the perihelion precession of Mercury, the recent experimental verification of the event horizon of a black hole, etc \cite{Eddington:1920ed,Das:2019khz,LaPlaca:2019rjz,Horizon:2010ap}.

All above experiments can be theoretically explained through of Schwarzschild's metric in which a spherically symmetric spacetime
is introduced. It was Karl Schwarzschild who first exactly solved Einstein's equation \cite{Schwarzschild:1916ks}. Several researches on general relativity takes Schwarzschild's spacetime in order to explain important aspects about the universe, as the possibility of the presence of negative mass matter in the world considering the mentioned metric as well as to better understand the dynamics of black holes \cite{Eddington:1920ed,Das:2019khz,LaPlaca:2019rjz, Bondarenko:2018ikv}. 

There are many important works aimed to investigate the solar system, in particular, related to the orbit of planets around the sun as well as the study of orbit corrections using general relativity instead of Newtonian mechanics \cite{Alexanian:2019xik,Shapiro:2002sh,Treu:2012tr}. Generally, the orbital characteristics of a particle moving around a source can be better investigated by the use of the geodesic equation in Schwarzschild's metric. The geodesic equation, commonly known as equation of motion, can be considered one of the useful tools to understand the dynamics of the system, and is one of the theoretical approaches capable of identifying the new effects introduced by the general relativity. In addition, perturbative methods are used to approximately solving the equations of motion, as in case of the orbit of Mercury, etc. Exact solutions of the orbit equations are mathematically complicated, and cannot be analytically solved in many situations.

There are some relevant exact solutions to Eq. 10 (see section \ref{section2}) in literature, taking distinct backgrouds \cite{Hackmann:2008zza, Hackmann:2008zz, Hackmann:2009nh, Hackmann:2010zz, Hackmann:2013pva}. A discussion on analytical methodologies to geodesic equations can be found in Ref. \cite{Hackmann:2015ewa}. The exact geodesics are also investigated in Ref. \cite{Scharf:2011ii} using Weierstrassian elliptic functions. A general treatment on elliptic integrals of geodesic equations in Schwarzschild's metric can be found in Ref. \cite{Chandrasekhar:BH}. Commonly are still presented approximative solutions in literature (see Ref. \cite{Padmanabhan:2010zzb}). It is important to emphasise that there is a broad discussion in Ref. \cite{Chandrasekhar:579245} about the geodesics of a massive particle to the cases in which $E>m$ and $E<m$, i. e., the source's energy is greater or less than its rest mass, respectively. These situations represent non-physical space-like geodesics when $E>m$, and the physical trajectories of the massive particle occurs for $E<m$ (see more details in Ref. \cite{Chandrasekhar:579245}). The restricted case in which $E=m$ need to be better investigated in order to verify the geometrical features of the geodesics. In add, it is interesting checking the physical aspects related to dynamic of the particle due to the exact solution.

In this paper, we are interested in provide an exact solution to geodesic equation in Schwarzschild backgroud, when the energy of system is  equal to rest mass of the massive particle, without consider any approximation or heuristic argument. These are important aspects addressed in this work about the geometry of the geodesics. Here, we explore exact geodesics in Schwarzschild's metric with a most transparent and simple form in terms of Jacobi's functions. It is considered Schwarzschild's spacetime to exact solving the orbit equation of a massive particle by using of Jacobi's elliptic functions. The obtained solution provides the relativistic effects on the massive particle that are absent in Newtonian gravity. What is more, we investigate the additional physical information introduced by the exact solution. We exactly solve the problem by the use an analytical methodology step by step in order to provide detailed solutions, presenting the particle trajectories in terms of Jacobis's elliptic functions with mathematical rigour.

This work is organized as follows: In section 1, we present a review on spherically symmetric spacetime. In section 2, as a complement of the previous section, we provide relevant aspects on the orbit equation in Schwarzschild's background as well as to prepare the orbit equation to be analytically solved in terms of Jacobi's elliptic functions. In section 3, we introduce an exact solution to the orbit equation. As it turns, we summarize our main findings and draw some perspectives. Throughout this work, we use units of $\hbar = c = 1.$

\qquad 
\section{A review on spherically symmetric spacetime}
\label{section2}
\qquad 
We know that Schwarzschild's metric is appropriate in physical problems that involves spherical symmetries, as in the study of the motion of planets around the sun, because it modifies the spacetime curvature. In particular, that metric can be useful to explain the perihelion of planets, the bending of light around the sun, etc. This approach describes the outside field of a spherically symmetric source, and the inside field in empty space \cite{Padmanabhan:2010zzb,Dienes:1998vh,Wheeler:1996ma,Rindler:1995qa}.

In order to derive an expression from Schwarzschild's metric is relevant emphasize some features of a spherically symmetric spacetime: the isotropy of spacetime with angular coordinates; there is non-rotation of the source; and the spacetime is static.    

The metric can be written as follows
\begin{eqnarray}
 ds^{2}&=&-A\left(r\right)dt^{2}+\frac{1}{A\left(r\right)}dr^{2}+\nonumber\\
        &+&r^{2}d\theta^{2}+r^{2}\sin^{2}\theta d\phi^{2} \qquad (r>R),
 \label{eq_1}
\end{eqnarray}
where $r$ is the distance from the source to an arbitrary position, $\theta$ is the inclination from the axis, and $\phi$ the azimuthal angle around the axis. In Eq. \ref{eq_1}, $R$ is known as Schwarzschild radii ($R=2GM$), and the term $A(r)$ is given by
\begin{equation}
A\left(r\right)=\left(1-\frac{2GM}{r}\right).
\label{eq_2}
\end{equation}

In next section, the above metric is used to determine the dynamics of the massive particle to obtain the orbit equation.

\section{Orbit equation in Schwarzschild's backgroud}
\label{section3}
\qquad
The Hamilton-Jacobi (H-J) relativistic equation (see Eq. \ref{eq_3}) describes more easily the motion of a particle of $m$ mass in Schwarzschild spacetime instead of Einstein's equation. Remembering that Einstein's equation is a second-order differential equation in spacetime coordinates and also a non-linear equation while Eq. \ref{eq_3} is a first-order differential equation \cite{Goldstein:1990aq,Arnold:1980ar} the H-J equation can be solved with less mathematical efforts.  

From the variational principle, we have know that given the action $\mathcal{S}\left(x^{\mu}\right)$ the energy and momentum of a particle of a massive are obtained in terms its derivatives $p_{\mu}=\partial S/\partial x^{\mu}$ and
$p_{\mu}p^{\mu}=-m^{2}$, and thus the H-J relativistic equation is given by
\begin{equation}
g^{\mu\nu}\frac{\partial S}{\partial x^{\mu}}\frac{\partial S}{\partial x^{\nu}}=-m^{2}.
\label{eq_3}
\end{equation}

The motion of the particle occurs only in the plane, i. e., $\theta =\pi /2$ and this emerges as consequence of the spherical symmetry. In Schwarzschild's metric the H-J equation becomes
\begin{equation}
 \frac{1}{A\left(r\right)}\left(\frac{\partial S}{\partial t}\right)^{2}
 -A\left(r\right)\left(\frac{\partial S}{\partial r}\right)^{2}
 -\frac{1}{r^{2}}\left(\frac{\partial S}{\partial\phi}\right)^{2}=-m^{2},
 \label{eq_4}
\end{equation}
where $E=(\partial S/\partial t)$ and $L=(\partial S/\partial\phi)$ are constants of motion, namely, energy and angular momentum. Besides, by substituting Eq. \ref{eq_2} in Eq. \ref{eq_4} it is possible to obtain 
\begin{eqnarray}
 \frac{E^{2}}{\left(1-2u\right)} +\frac{u^{4}\left(1-2u\right)}{G^{2}M^{2}}\left(\frac{dS}{du}\right)^{2}
 \nonumber\\
 -\frac{u^{2}L^{2}}{G^{2}M^{2}}=-m^{2}.&&
 \label{eq_5}
\end{eqnarray}
Making the variable change $u=GM/r$ in Eq. \ref{eq_5} and after applying the variable separation method, we have
\begin{eqnarray}
\left(\frac{dS}{du}\right)^{2}=-\frac{G^{2}M^{2}m^{2}}{u^{4}\left(1-2u\right)}+
\frac{G^{2}M^{2}E^{2}}{u^{4}\left(1-2u\right)}-\frac{L^{2}}{u^{2}\left(1-2u\right)}\\
=\frac{G^{2}M^{2}}{u^{4}(1-2u)}\left(E^{2}-m^{2}\right) -\frac{L^{2}}{u^{2}(1-2u)}.
\label{eq_6}
\end{eqnarray}
Solving Eq. \ref{eq_6} by integration it is possible to find  
 \begin{eqnarray} 
S=-Et+L\phi + f\left(u\right),
\label{eq_7}
\end{eqnarray}
where
\begin{eqnarray}
f\left(u\right)=\int \left[\frac{G^{2}M^{2}}{u^{4}(1-2u)}\left(E^{2}-m^{2}\right) -\frac{L^{2}}{u^{2}(1-2u)}\right]^{\frac{1}{2}}du.
\label{eq_8}
\end{eqnarray}
It is important to observe that appears in Eq. \ref{eq_7} the energy $E$, the angular momentum $L$, and the function $f\left(u\right)$ defined by Eq. \ref{eq_8} that represents the geodesic of the particle. Here, we are interested in solving the integral given by Eq. \ref{eq_8} that provides the behavior of the massive particle, and particularly its geodesic equations. 

In order to determine the geodesic equations it is necessary taking $\partial S/\partial L =0$ in Eq. \ref{eq_7} that leads the integral equation below
\begin{eqnarray}
\phi=\nonumber\\
\int\frac{Ldu}{\sqrt{-\left(1-2u\right)G^{2}M^{2}m^{2}-L^{2}u^{2}\left(1-2u\right)+G^{2}M^{2}E^{2}}}.
\label{eq_9}
\end{eqnarray}

In more general studies about geodesic equations Eq. \ref{eq_9} can be used to investigate both cases where $E=m$ and $E\neq m$ \cite{Chandrasekhar:579245}. However, generally are presented only numerical or non-anallytical solutions to the orbits \cite{Padmanabhan:2010zzb,Dienes:1998vh}. Here, as mentioned in introduction, we solved the case $E=m$ using an analytical methodology whereby are presented the exact orbits of the particle.

It is important to emphasize that there are some exact solutions in distinct backgrouds to Eq. \ref{eq_9} \cite{Hackmann:2008zza, Hackmann:2008zz, Hackmann:2009nh, Hackmann:2010zz, Hackmann:2013pva}. However, we introduce in this paper a relevant review about the geodesic equations of a massive particle in Schwarzschild background, wherby the orbit equations are exactly solved using an analytical methodology in terms of Jacobi's elliptic functions. The step by step development is by itself extremely useful to understand important mathematical aspects associated to the solutions of geodesic equations. An exact solution in terms of those functions is presented in the next section, the core of this work.

\qquad
\section{An exact solution in terms of Jacobi's elliptic functions}
\label{section4}
\qquad

Rewritting Eq. \ref{eq_9} in its differential form
\begin{eqnarray}
\frac{du}{d\phi}=\nonumber\\
\frac{1}{L}\sqrt{-\left(1-2u\right)G^{2}M^{2}m^{2}-L^{2}u^{2}\left(1-2u\right)+G^{2}M^{2}E^{2}},
\label{eq_10}
\end{eqnarray}
and after some simple algebraic manipulations in Eq. \ref{eq_10} and considering the restriction addressed in this work or special case in which $E=m$, and taking $\kappa=GMm/L$ it is possible to define
\begin{equation}
g\left(u\right)=2u^{3}-u^{2}+2\kappa^{2}u,
\label{eq_11}
\end{equation}
where $g\left(u\right)$ is the integrated function on the right side in Eq. \ref{eq_8}.

A polynomial form of Eq. \ref{eq_11} in terms its roots and conveniently written to be analytically solved is given by
\begin{eqnarray}
\left(\frac{du}{d\phi}\right)^{2}=f\left(u\right)=
2\left(u-u_{1}\right)\left(u-u_{2}\right)\left(u-u_{3}\right),
\label{eq_12}
\end{eqnarray} 
where $u_{1}$, $u_{2}$ and $u_{3}$ are obtained solving the equation $2u^{3}-u^{2}+2\kappa^{2}u=0$, in which the roots are written below
\begin{eqnarray}
u_{1}&=&\frac{1}{4}-\frac{1}{4}\sqrt{1-16\kappa^{2}},\nonumber\\
u_{2}&=&0,\nonumber \\
u_{3}&=&\frac{1}{4}+\frac{1}{4}\sqrt{1-16\kappa^{2}}.
\label{eq_13}
\end{eqnarray}
The equation given by expression \ref{eq_12} governing the orbit of a massive particle in Schwarzschild's metric and can be solved using Jacobi's elliptic functions, as follows described.

Let us start with a brief review on Jacobi elliptic functions in Legendre's form that are useful to exactly solve Eq. \ref{eq_12} \cite{Abramowitz:1970ma}. 
The elliptic integral that leads to Jacobi functions in Legendre's form is shown below
\begin{equation}
x=\int_{0}^{\theta}\frac{d\psi}{\sqrt{1-\beta\sin^{2}(\psi)}}.
\label{eq_14}
\end{equation}
In the expression given by Eq. \ref{eq_14} $\beta\in(-1,1)$ is namely modulus. Besides this, the square root is a positive value. Defining the following coordinate transformations
\begin{equation}
y=\sin(\alpha)\equiv\mbox{sn}[x|\beta],
\label{eq_18}
\end{equation}
the notation $\mbox{sn}[x|\beta]$ is commonly used to denote the known Jacobi's elliptic functions.

See that the integral form of Eq. \ref{eq_12} written as (see more details in Ref. \cite{Abramowitz:1970ma})
\begin{equation}
\phi(u)=\int\frac{du}{\sqrt{2\left(u-u_{1}\right)\left(u-u_{2}\right)\left(u-u_{3}\right)}},
\label{eq_15}
\end{equation}
can be compared with Eq. \ref{eq_14} by the appropriate coordinate transformations. Another way of comparing the orbit equation obtained here (see Eq. \ref{eq_12}) with the expression of the elliptic integral given by Eq. \ref{eq_14}, is rewriting the latter in a differential form, as presented follows. It is interesting to stress that, any integrated expression containing a third or fourth degree polynomial in the denominator of a fraction can be reduced to an elliptic integral \cite{Abramowitz:1970ma}.

Resuming the previous discussion and deriving Eq. \ref{eq_14}
\begin{equation}
\frac{dx}{d\theta}=\frac{1}{\sqrt{1-\beta\sin^{2}(\theta)}},
\label{eq_16}
\end{equation}
it is possible to obtain
\begin{equation}
\left(\frac{d\theta}{dx}\right)^{2}=1-\beta\sin^{2}(\theta).
\label{eq_17}
\end{equation}
Combining Eqs. \ref{eq_17} and \ref{eq_18}, we obtain
\begin{equation}
\left(\frac{d\theta}{dx}\right)^{2}=1-\beta y^{2}.
\label{eq_19}
\end{equation}

By the chain rule, we have
\begin{equation}
\frac{dy}{dx}=\frac{dy}{d\theta}\frac{d\theta}{dx}=\cos(\theta)\frac{d\theta}{dx}
\label{eq_20}
\end{equation}
that provides
\begin{equation}
\left(\frac{dy}{dx}\right)^{2}=\left(1-y^{2}\right)\left(1-\beta y^{2}\right).
\label{eq_21}
\end{equation}

We need to write the above expression the same way as the orbit equation of a massive particle given by Eq. \ref{eq_12}. To do that, we take the coordinate transformation and by substituting Eq. \ref{eq_22} in Eq. \ref{eq_21}
\begin{equation}
y^{2}=az+b
\label{eq_22}
\end{equation}
in which provides after some simple algebraic procedures Eq. \ref{eq_23}
\begin{equation}
\left(\frac{dy}{dx}\right)^{2}=\frac{a^{2}}{4y^{2}}\left(\frac{dz}{dx}\right)^{2},
\label{eq_23}
\end{equation}
where $dz/dx$ in Eq. \ref{eq_23} is given by
\begin{equation}
\left(\frac{dz}{dx}\right)^{2}=\frac{4}{a^{2}}\left(az+b\right)
\left(az+b-1 \right)\left[\beta(az+b)-1 \right].
\label{eq_24}
\end{equation}

In order to rewrite this equation similarly to Eq. \ref{eq_12}, we need to take the transformations below, making $z=u$, $x=\alpha\phi$, and so
\begin{eqnarray}
\left(\frac{du}{d\phi}\right)^{2}&=&\frac{4\alpha^{2}}{a^{2}}\left(au+b\right)\nonumber\\
&\times&\left(au+b-1 \right)\left[\beta(au+b)-1 \right],
\label{eq_25}
\end{eqnarray}
whereby we obtain
\begin{eqnarray}
\left(\frac{du}{d\phi}\right)^{2}&=&4\alpha^{2}\beta a\left(u+\frac{b}{a}\right)\nonumber\\
&\times&\left(u+\frac{b-1}{a} \right)\left(u+ \frac{b}{a}-\frac{1}{a\beta} \right).
\label{eq_26}
\end{eqnarray}

Note that Eq. \ref{eq_26} already has the desired aspect and can be compared with Eq. \ref{eq_12} to obtain $u_1$, $u_2$ and $u_3$, as follows
\begin{eqnarray}
4\alpha^{2}\beta a\left(u+\frac{b}{a}\right)\left(u+\frac{b-1}{a} \right)\left(u+ \frac{b}{a}-\frac{1}{a\beta} \right)&\nonumber\\
=2\left(u-u_{1}\right)\left(u-u_{2}\right)\left(u-u_{3}\right)&.
\label{eq_27}
\end{eqnarray}

Observe that the equality given by Eq. \ref{eq_27} provides
\begin{eqnarray}
u_{1}=-\frac{b}{a},\quad u_{2}=\frac{1}{a}+u_{1},\quad u_{3}=\frac{1}{a\beta}+u_{1},
\label{eq_28}
\end{eqnarray}
and
\begin{eqnarray}
\alpha=\sqrt{\frac{u_{3}-u_{1}}{2}},\quad \beta=\frac{u_{2}-u_{3}}{u_{3}-u_{1}}.
\label{eq_29}
\end{eqnarray}
Taking the values of the Eqs. \ref{eq_28} and \ref{eq_29} in Eq. \ref{eq_22}, we have
\begin{eqnarray}
u=u_{1}+\left(u_{2}-u_{1}\right)\mbox{sn}^{2}[\alpha\phi |\beta].
\label{eq_30}
\end{eqnarray}

It is worthy of emphasis that Eq. \ref{eq_30} represents the exact solution of the orbit equation (see Eq. \ref{eq_12}) of a massive particle in Schwarzschild's metric in terms of Jacobi's elliptic functions. As $u=GM/r$, we obtain the final orbit equation. Notice that this expression is derived without any approximation, being an exact solution that governing the motion of a massive particle in Schwarzschild's metric.
\begin{eqnarray}
\frac{1}{r}=\frac{1}{r_{1}}+\left(\frac{1}{r_{2}}-\frac{1}{r_{1}}\right)\mbox{sn}^{2}[\alpha\phi |\beta].
\label{eq_31}
\end{eqnarray}

In order to investigate some physical features of the trajectory of a massive particle we need to plot Eq. \ref{eq_30} (see also the values of $u_i$(i=1, 2,3) in Eq. \ref{eq_13}), considering the orbital angle versus the distance of the particle to the source. Thus, $-1/4\leq\kappa\leq 1/4$. As consequence, the geodesic of the particle should be analyzed in different regimes, where the real orbits physically occurs only if the root squares in Eq. \ref{eq_13} were always positive. Besides, the well defined $k$ value imposes a constraint to the particle's geodesic, that is, for $k=1/4$ we have the non-relativitic orbit of the massive particle, and the other ranges can be seen in figures below. For the Fig. \ref{fig0}, we take $u_{1}$ and $u_{3}$ positives, and it is possible to observe a periodic behaviour of the orbit.
\begin{figure}[!htbp]
	\begin{center}
		\includegraphics[width=0.4\textwidth]{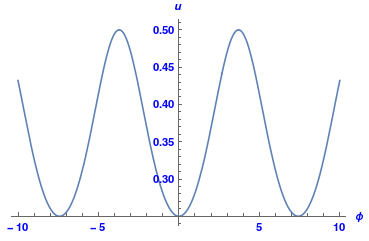}
		\caption{Orbital angle versus distance from the particle to the source. $G=1$. A periodic behaviour of the orbit is observed. Besides, $u_{1}$ and $u_{3}> 0$.}\label{fig0}
	\end{center}
\end{figure}
If we take the $u_{1}=0$ and $u_{3}=1/4$, see in Fig. \ref{fig1} an oscillatory motion with shortened period.

\begin{figure}[!htbp]
	\begin{center}
		\includegraphics[width=0.4\textwidth]{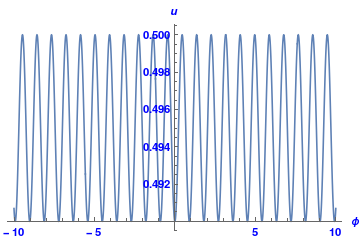}
		\caption{Orbital angle versus distance from the particle to the source. $G=1$. A shortened period of the orbit is observed. Besides, $u_{1}=0$ and $u_{3}> 1/4$.}\label{fig1}
	\end{center}
\end{figure}
\begin{figure}[!htbp]
	\begin{center}
		\includegraphics[width=0.4\textwidth]{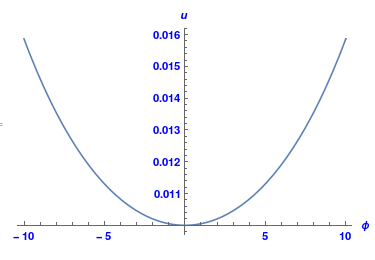}
		\caption{Orbital angle versus distance from the particle to the source. $G=1$. A non-periodic behaviour of the orbit is observed with bounded orbits, and the particle is trapped in gravitational field of the source. Besides, we take $u_{1}<0$ and $u_{3}>0$.}\label{fig2}
	\end{center}
\end{figure}

For the Fig. \ref{fig2}, we take $u_{1}<0$ and $u_{3}>0$, and it is possible to observe a non-periodic behaviour of the geodesic, showing bounded orbits in which the particle is trapped in the gravitational field of the source.
\qquad 
\section{Concluding remarks}
\label{section5}
\qquad
We show that Jacobi's elliptic functions are appropriate to exactly solve an elegant way the orbit equation of a massive particle in  Schwarzschild's metric.

We provide an analytical solution that captures the relativistic effects on the massive particle, and investigate the additional physical information introduced by the exact solution as well as the geometry of the geodesics. 

We find oscillatory movements of the orbit of the massive particle  without to consider heuristic arguments as well as a region where the massive particle shows a confinement around the source. 

We hope that the present work can stimulate further contributions using others backgrounds, and can be used of protocol in order to investigate the possibility of exactly solve geodesic equations to charged particle case. Besides, the same methodology is naturally useful to study massless particles.

\qquad
\section{Acknowledgments}
\qquad
The authors gratefully acknowledge the support provided by Brazilian Agencies CAPES, CNPQ. We would like to thank the following for their kind support: Piau\'i Federal Institute, S\~ao Raimundo Nonato campus; for our friend and colleague Israel A. C. Noletto, for briefly proofreading this paper. 

\bibliographystyle{unsrt}  


\bibliographystyle{elsarticle-num}
\bibliography{references}

\end{document}